# Compact QED in Landau Gauge: a lattice gauge fixing case study

M.I. Polikarpov
*ITEP, Moscow, 117259, Russia*
e-mail: polykarp@vxdsyc.desy.de

Ken Yee
*Dept. of Physics and Astronomy, L.S.U.*
*Baton Rouge, Louisiana   70803-4001, USA*
e-mail: kyee@rouge.phys.lsu.edu

M.A. Zubkov
*ITEP, Moscow, 117259, Russia*
e-mail: zubkov@vxitep.itep.msk.su

May 24, 1993

**Abstract**

We derive different representations of compact QED fixed to Landau gauge by the lattice Faddeev-Popov procedure. Our analysis finds that (A)Nielsen-Olesen vortices arising from the compactness of the gauge-fixing action are *quenched*, that is, the Faddeev-Popov determinant cancels them out and they do not influence correlation functions such as the photon propagator; (B)Dirac strings are responsible for the nonzero mass pole of the photon propagator. Since in $D = 3 + 1$ the photon mass undergoes a rapid drop to zero at $\beta_c$, the deconfinement point, this result predicts that Dirac strings must be sufficiently dilute at $\beta > \beta_c$. Indeed, numerical simulations reveal that the string density undergoes a rapid drop to near zero at $\beta \sim \beta_c$.

# 1 Introduction and Results

Gauge fixing is essential to several potentially physically relevant lattice computations, of which we mention two: (i)Partial gauge fixing of $SU(3)$ to residual $U(1) \times U(1)$ is necessary to define abelian projection monopoles [1, 2] whose currents, as a working hypothesis, may be the underlying confinement mechanism of QCD. (ii)Computing fixed-gauge lattice matrix elements may be a way to determine continuum-lattice renormalization or "matching" coefficients. In particular, such coefficients are necessary (in a certain approach) for trying to exhibit the Delta I=1/2 Rule on the lattice [3].

With such ultimate motivations, there have been many numerical studies of lattice gauge fixing and the gauge dependence of such gauge variant quantities as the abelian projection monopole density [1, 2, 4] and effective gluon, quark, photon and electron masses [5, 6, 7]. In addition, the gauge dependence of quark masses has been analytically computed for certain gauges in the strong-coupling expansion [8] and in the Schwinger model [9].

While compact QED("cQED") in the strong coupling regime in the absence of gauge fixing is a well-studied model of a confining gauge theory [11], its fixed-gauge features are less known and also nontrivial [6, 12]. In this paper, we report on a numerical and analytical study of cQED fixed to Landau gauge. As described in Section 2 *both* the photon mass[1] and Dirac string density("kink" density) drop dramatically from nonzero to near zero at $\beta \sim \beta_c$, the deconfinement point in $D = 3+1$ dimensions. Is this a coincidence or are Dirac strings dynamically related to photon mass? Sections 3 and 4 cast the

---
[1]We stress that photon "mass" in this paper refers to the pole of the photon propagator. Photon mass thusly defined is gauge variant and is *not* obviously related to such physical length scales as the electric penetration depth, which is gauge invariant.

lattice Faddeev-Popov procedure into lattice differential forms notation [13], using which we show that:

- (A)Abrikosov-Nielsen-Olesen [15] vortices from the lattice gauge fixing "spin glass" action are quenched and, so, do not contribute to photon mass;

- (B)Disorder caused by Dirac strings is responsible for photon mass [12].

## 2 Numerical Results

The cQED action is

$$S_c \equiv \beta \sum_{\mu<\nu}(1-\cos F_{\mu\nu}) \tag{1}$$

where $F_{\mu\nu}[A] \equiv \partial_\mu A_\nu - \partial_\nu A_\mu$ and $A_\mu \in [-\pi,\pi)$. While $S_c$ is gauge invariant,

$$F_{\mu\nu} \equiv \Theta_{\mu\nu} + 2\pi N_{\mu\nu} \tag{2}$$

decomposes into a gauge invariant electromagnetic field part $\Theta_{\mu\nu} \in [-\pi,\pi)$ and an integral part $N_{\mu\nu}$. If $N_{\mu\nu} \neq 0$ on a plaquette, we say that the plaquette has a "kink." Dirac strings, which costs zero action, and magnetic monopoles, the Dirac string endpoints which cost action, are comprised of kinks. Since the condition $A_\mu \in [-\pi,\pi)$ is enforced by $2\pi$ modding, $N_{\mu\nu}$ transforms nontrivially under local gauge transformations which push $A_\mu$ outside of $[-\pi,\pi)$ and, so, Dirac strings are gauge variant. Monopoles are gauge invariant.

Landau gauge in this paper is defined as in Ref. [6]. As described in Appendix A, an alternative definition—henceforth called "cLandau" gauge—was adopted in Ref. [12]. cLandau gauge is a compactness-preserving gauge which can heuristically be thought of as $(\partial_\mu A_\mu) \mod(2\pi) = 0$. In the terminology of this paper, the "Landau" gauge photon mass values reported in [12] are really in cLandau (not Landau) gauge. Photon masses quoted in this paper are in Landau gauge.

While the gauge variant Dirac strings do not have any direct physical effects, they play a central role in the fixed-gauge sector. Figure 1 depicts a typical Landau gauge configuration in cQED$_{2+1}$ on a $25^3$ lattice with periodic boundary conditions. The thin lines mark the Dirac strings, the big dots the monopoles and antimonopoles. The reader should be convinced that Dirac strings either connect monopole-antimonopoles pairs or form closed loops. In Landau gauge the links tend to $U_{x,\mu} \to 1$ at large $\beta$ and at $\beta = 2.3$ the operator

$$F(x) \equiv \frac{1}{2} \sum_{\mu=1}^{3} [\cos A_\mu(x) + \cos A_\mu(x - \hat{\mu})] \tag{3}$$

has a typical value of $F(x) \sim 0.95$. However, $3\%$ of the sites obey $F(x) \leq 0.77$. These "small-F" sites are indicated in Figure 1 by the small dots.

Let us mention that we have seen analogous small-F sites in the course of gauge fixing $SU(3)$ gauge theory to maximal abelian gauge and Landau gauge. While we do not claim to understand at this point that small-F sites of cQED have anything in common with QCD small-F sites, we have determined that $SU(3)$ small-F sites have (quite dramatically) fractal dimension $D_f \sim 2$ for Landau gauge and, less dramatically, a bit smaller $D_f$ for maximal abelian gauge. These $SU(3)$ results are preliminary and will not be further discussed here [2].

In Figure 1 the small-F sites cluster around closed and open Dirac strings because the vector potential around a Dirac string in the $\hat{z}$ direction is (in continuum cylindrical coordinates) $\vec{A}^{\text{string}} = \hat{\phi} \frac{1}{\sqrt{x^2+y^2}}$ and around a monopole at $\vec{x} = 0$ is (in spherical coordinates) $\vec{A}^{\text{mono}} = \hat{\phi} \frac{1-\cos\theta}{r\sin\theta}$. Consequently, near Dirac strings and monopoles the smoothness of the vector potential is disrupted. *This disruption disorders the fixed-gauge sector of the cQED vac-*

Figure 1: A typical $\beta = 2.3$ cQED$_{2+1}$ gauge configuration on a $25^3$ periodic lattice after about $\sim 2000$ 30-hit Metropolis/pseudo-heatbath thermalization sweeps. The configuration is fixed to Landau gauge. The thin lines mark the Dirac strings, the bigger dots the magnetic monopoles and antimonopoles, and the smaller dots the locations of the smallest 3% of $F(x)$.

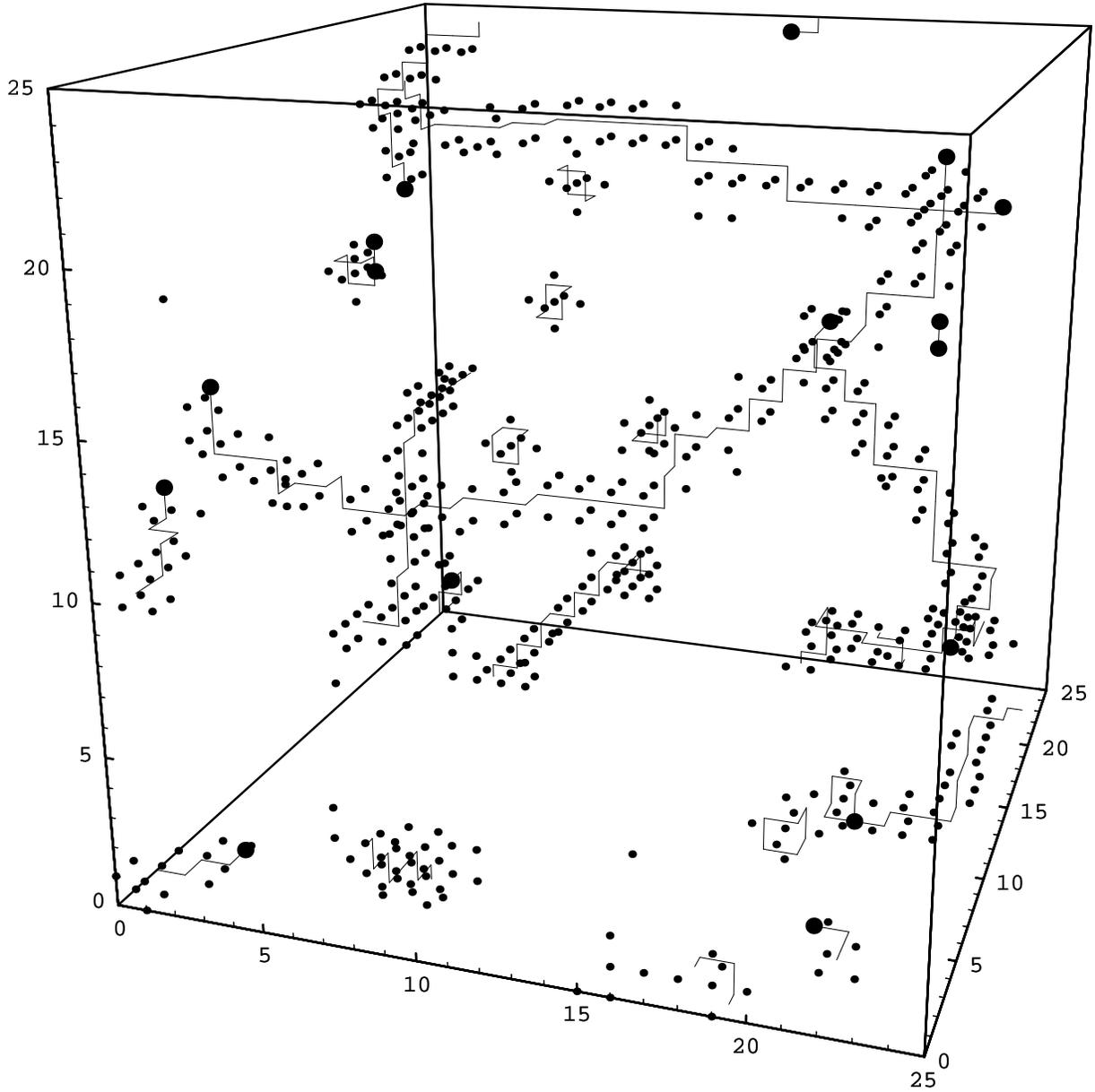

*uum*. As described in Section 4, one consequence of this disorder is that the mass pole $M_\gamma$ of the photon propagator is nontrivial.

Figure 2 depicts the Landau gauge kink number density

$$\rho_N \equiv \frac{\#\text{kinks in the lattice}}{\#\text{plaquettes in the lattice}} \qquad (4)$$

as a function of $\beta$ in $D = 3 + 1$ dimensions. Since $N_{\mu\nu}$ is gauge variant, $\rho_N$ varies with gauge. In Landau gauge, like the photon mass reported in Ref. [6] $\rho_N$ drops dramatically to near zero at $\beta \sim \beta_c$. Figures 3 and 4 show that photon mass (when it is nonzero) is, approximately, linearly correlated with $\rho_N$ in both $D = 3 + 1$ and $D = 2 + 1$ dimensions. Note that in $D = 2 + 1$ we are quoting the dimensionless quantity $\beta M_\gamma$ rather than $M_\gamma$. The lines in Figures 3 and 4 are to guide-the-eye.

In Landau gauge $\rho_N$ and monopole density $\rho_M$ are positively correlated. ($\rho_M$ also drops sharply near $\beta_c$ [16].) This is because Landau gauge suppresses Dirac string length—most of the contribution to $\rho_N$ comes from unavoidable Dirac strings which connect monopoles-antimonopole pairs. Moreover, while monopole-prohibition (the procedure described in Ref. [12]) decreases both $\rho_N$ and $M_\gamma$ substantially in Landau gauge, $\rho_N$ and $M_\gamma$ decrease together so that they remain on the original $\rho_N$-$M_\gamma$ line (see Figure 4).

On the other hand, since $(\partial_\mu A_\mu)\text{mod}(2\pi) = 0$ leaves $\pm\pi$ ambiguities on $A_\mu$, cLandau gauge does not suppress Dirac strings. In this gauge monopole prohibition does not decrease kink density much—most of the kinks are in closed Dirac strings. Hence in cLandau gauge monopole prohibition does not significantly change the photon mass [12]. This is another piece of evidence linking photon mass to Dirac string density.

In summary, Landau gauge suppresses Dirac strings whereas other gauges such as cLandau do not. In any gauge and in both $D = 2+1$ and $D = 3+1$, when Dirac strings are dilute the photon mass is small; as Dirac strings be-come denser, the photon mass becomes larger. This monotonic relationship holds whether monopoles are prohibited or not.

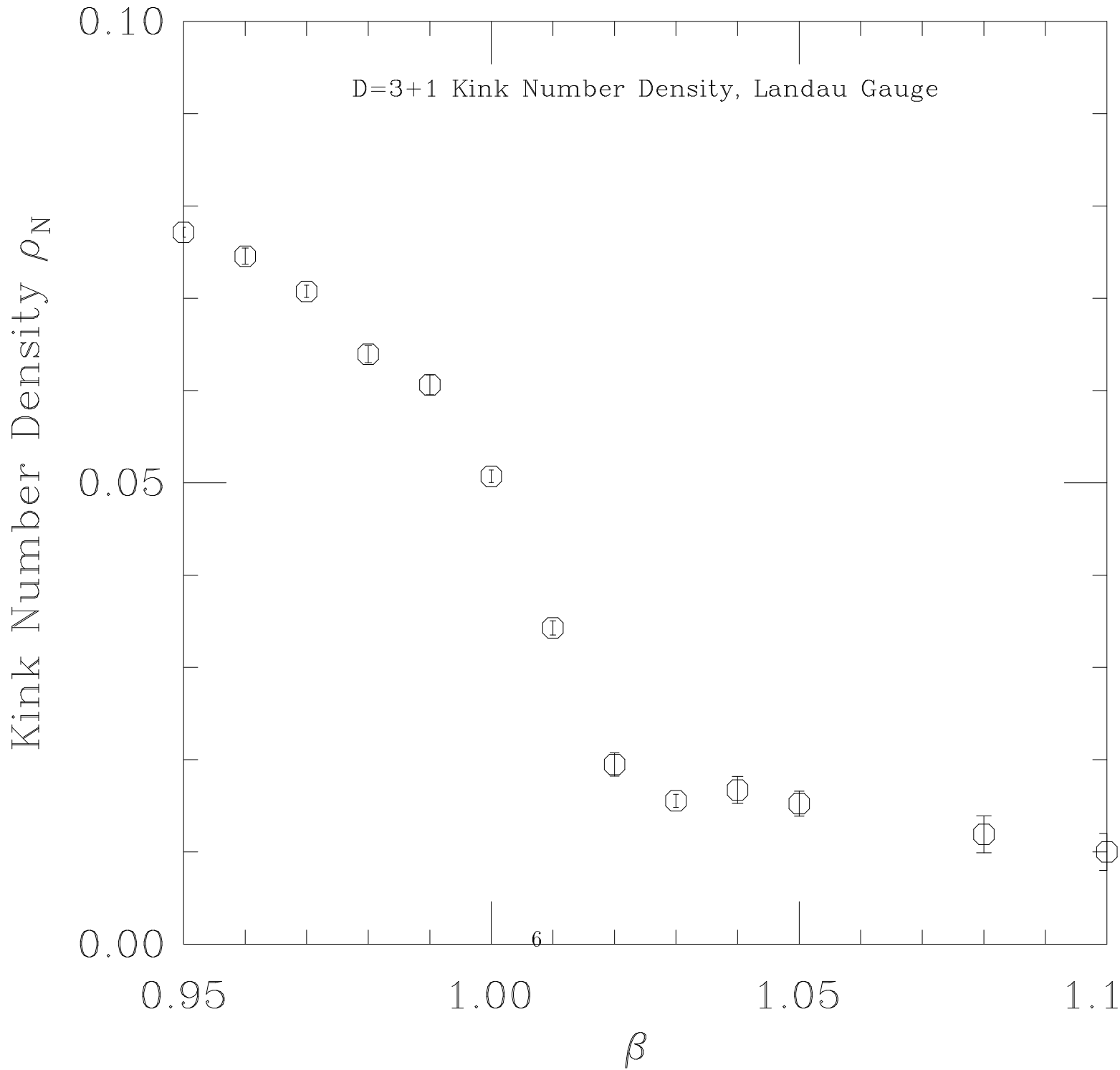

Figure 2: The Landau gauge kink number density $\rho_N$ as a function of $\beta$ in cQED$_{3+1}$ on an $8^3 \times 16$ periodic lattice. The first configurations at each $\beta$ is thermalized starting from cold-start by 2000 30-hit Metropolis/pseudo-heatbath sweeps. The error bars indicate the standard deviation of $\rho_N$ in 20 successive gauge configurations separated by 40 sweeps.

Figure 3: A scatter plot showing the correlation between the Landau gauge photon mass $M_\gamma$ and the Landau gauge kink number density $\rho_N$ at common $\beta$s in cQED$_{3+1}$ on an $8^3 \times 16$ lattice. The gauge configuration generation is done as for Figure 2 except that in this case 500 gauge configurations separated by 5 sweeps were used for each data point. We stress that each data point in this Figure is generated fresh from a cold-start *independently* of any other data point so there is statistical correlation between the plotted points. Errors are jackknife errors. The line is to guide-the-eye.

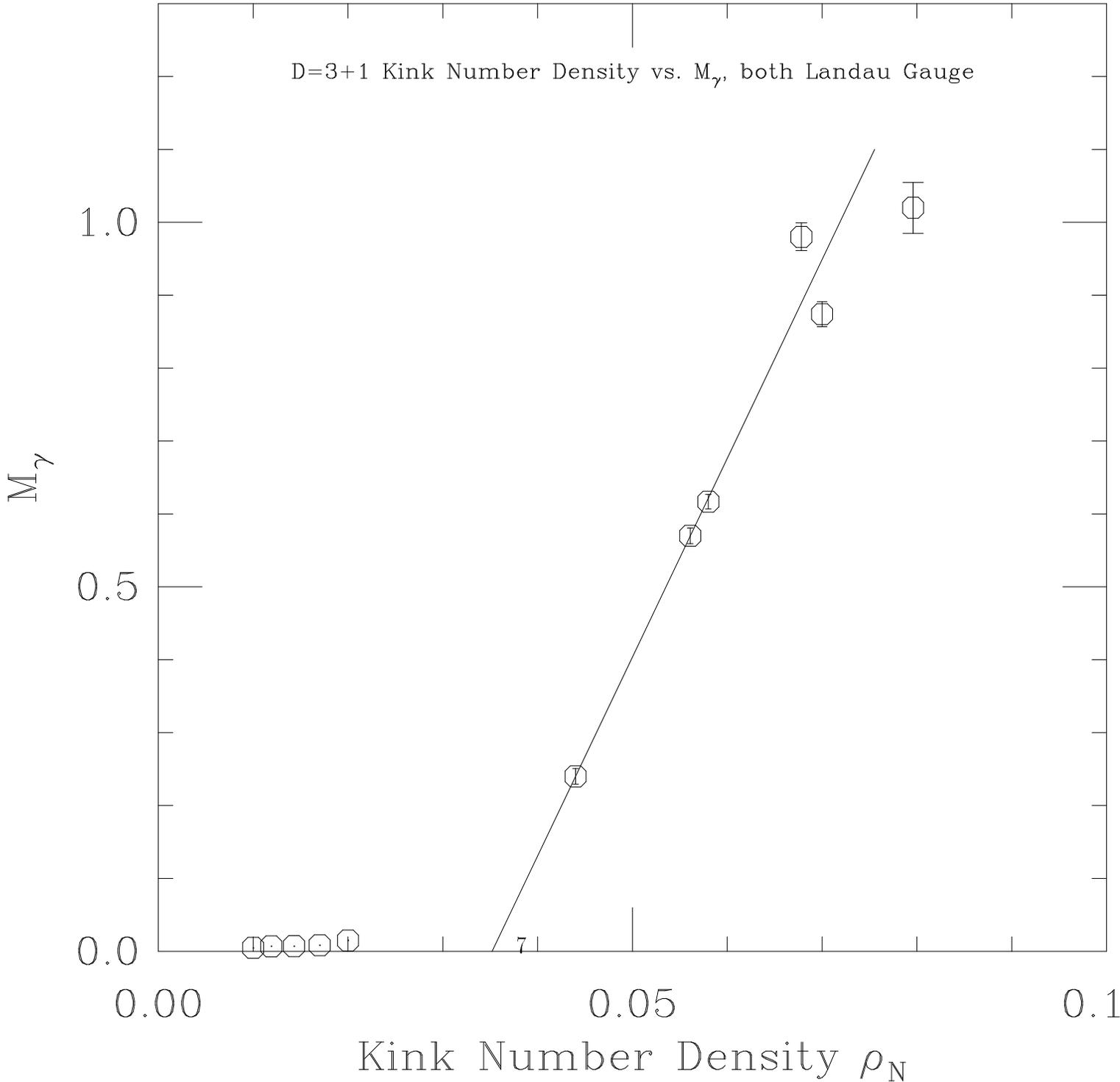

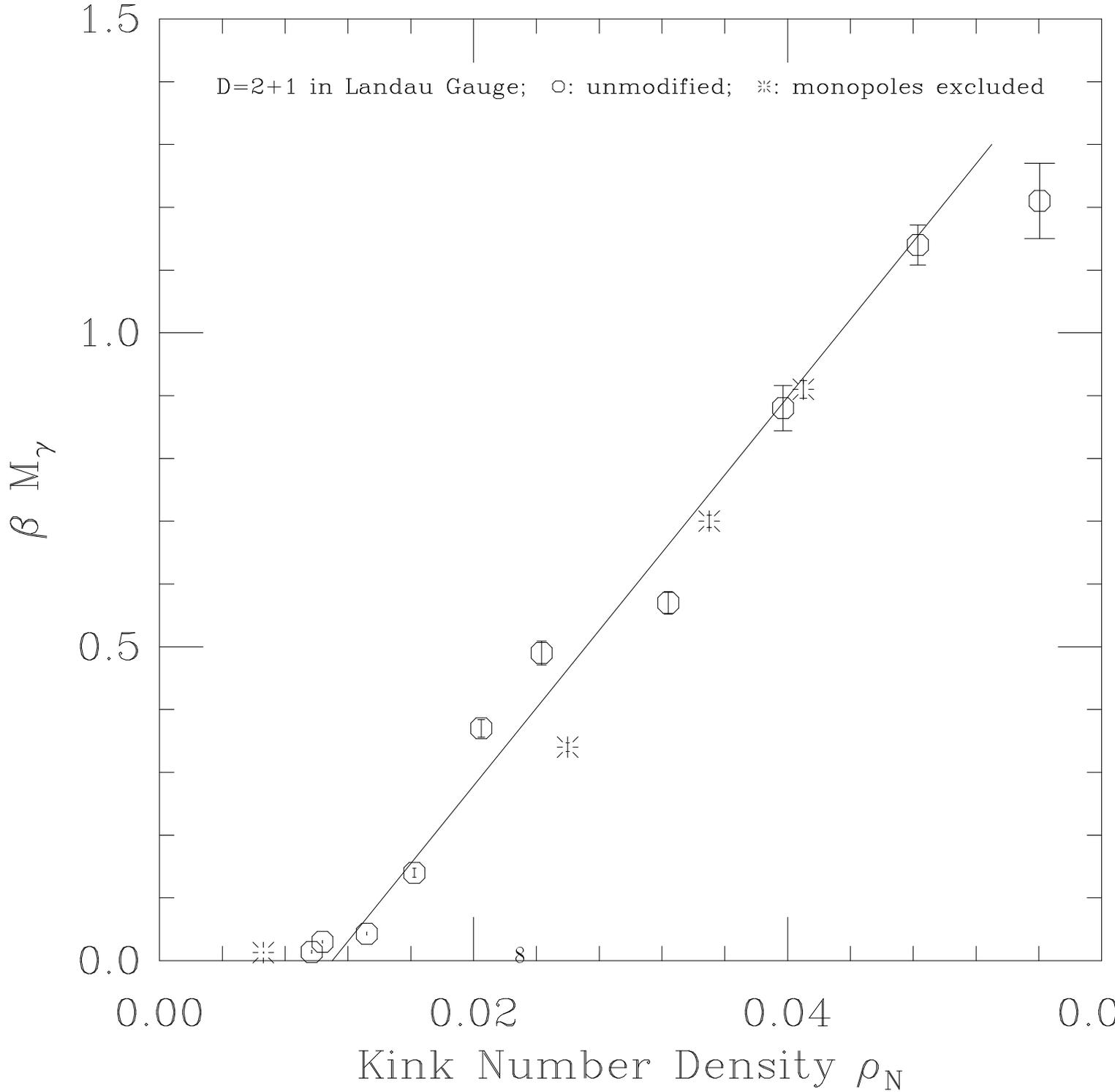

Figure 4: Same as Figure 3 except (i) we plot $\beta \times M_\gamma$ rather than $M_\gamma$; (ii) the calculation is in cQED$_{2+1}$ rather than in cQED$_{3+1}$; (iii) we include data points (the asterisks) for which monopoles are prohibited during the Metropolis/pseudo-heatbath updating sweeps.

## 3  Quenching of Spin Glass Vortices

In this Section we apply lattice differential form and BKT [10] transformation ideas, previously exploited to expose cQED with*out* gauge fixing [11] and the compact abelian Higgs model [13], to analyze the lattice Faddeev-Popov determinant arising in lattice gauge fixing. The strategy is to perform a variables change so that the underlying excitations—spin glass vortices arising from the periodicity of the lattice gauge fixing action—are explicitly displayed. We will show that these vortices are quenched by the Faddeev-Popov determinant and, so, do not play an important role in disordering such gauge variant correlation functions as the photon propagator. These spin glass vortices should *not* be confused with Dirac strings, which come from the cQED action.

The gauge fixed expectation value of a gauge variant operator $\mathcal{O}(A) \neq \mathcal{O}(A - \mathrm{d}\varphi)$ as computed in numerical simulations is

$$\langle \mathcal{O}(A_{\text{gauge}}^{\text{fixed}}) \rangle \equiv \frac{\int \mathcal{D}A \ \exp\{-S_c(A)\} \ \mathcal{O}(A_{\text{gauge}}^{\text{fixed}})}{\int \mathcal{D}A' \ \exp\{-S_c(A')\}} \tag{5}$$

where path integral measure $\mathcal{D}A$ is not gauge-fixed. Definition (5) is equivalent to the Faddeev-Popov approach since, with a suitable choice of gauge fixing action $S_{\text{gf}}(A) \neq S_{\text{gf}}(A - \mathrm{d}\varphi)$, operator $\mathcal{O}(A_{\text{gauge}}^{\text{fixed}})$ can be represented as

$$\mathcal{O}(A_{\text{gauge}}^{\text{fixed}}) = \Delta_{FP}(A) \int \mathcal{D}\varphi \ \exp\{-S_{\text{gf}}(A - d\varphi)\} \mathcal{O}(A - \mathrm{d}\varphi). \tag{6}$$

The substitution $\mathcal{O} = 1$ into (6) yields for the inverse Faddeev-Popov determinant

$$\Delta_{FP}^{-1}(A) = \int_{-\pi}^{\pi} \mathcal{D}\varphi \ \exp\{-S_{\text{gf}}(A - d\varphi)\}. \tag{7}$$

Note that $\Delta_{FP}(A) = \Delta_{FP}(A - \mathrm{d}\varphi)$. Hence as in continuum models the Faddeev-Popov determinant is gauge invariant. Inserting (6) into (5) and using the gauge invariance of $S_c$ and $\Delta_{FP}$ to transform away the $\varphi$-dependence of the $\mathcal{D}A$ integrand yields [7, 8]

$$\langle \mathcal{O}(A_{\substack{\text{fixed} \\ \text{gauge}}}) \rangle = \mathcal{Z}[\mathcal{O}]/\mathcal{Z}[1] , \qquad (8)$$

$$\mathcal{Z}[\mathcal{O}] \equiv \int \mathcal{D}A \, \exp\{-S_c(A)\} \, G(A) \, \mathcal{O}(A), \qquad (9)$$

$$G(A) \equiv \Delta_{FP}(A) \, \exp\{-S_{\mathrm{gf}}(A)\} . \qquad (10)$$

While our numerical simulations are performed in the Wilson representation, we will now adopt the Villain representation, which is more amenable to formal analysis. There is no doubt that for our purposes the Wilson and Villain approaches are qualitatively (albeit not quantitatively) equivalent. In the Villain representation

$$S_c(A) \equiv -\log \sum_{N(\mathrm{c}_2) \in \mathbb{Z}} \exp\left\{-\frac{\beta}{2}\|\mathrm{d}A - 2\pi N\|^2\right\}, \qquad (11)$$

$$S_{\mathrm{gf}}(A) \equiv -\log \sum_{l(\mathrm{c}_1) \in \mathbb{Z}} \exp\left\{-\frac{\kappa}{2}\|A - 2\pi l\|^2\right\} . \qquad (12)$$

The $\kappa \to \infty$ limit corresponds to minimizing $S^L$ of Eq. (A.1). Following from (7) and (12), the inverse Faddeev-Popov determinant is

$$\Delta_{FP}^{-1}(A) = \sum_{l(\mathrm{c}_1) \in \mathbb{Z}} \int_{-\pi}^{\pi} \mathcal{D}\varphi \, \exp\left\{-\frac{\kappa}{2}\|A - \mathrm{d}\varphi - 2\pi l\|^2\right\} . \qquad (13)$$

In (13) $\Delta_{FP}^{-1}$ is equivalent to the partition function of the compact abelian Higgs model in an external electromagnetic field.

One can manipulate lattice differential forms just like continuum ones. Our integration by parts convention is $(\varphi, \delta\psi) = +(\mathrm{d}\varphi, \psi)$ so that $\delta \equiv + *\mathrm{d}*$ in any dimensions [13]. The Laplacian is $\Delta \equiv \delta\mathrm{d} + \mathrm{d}\delta$ and $j$-forms are generically denoted $c_j$. Zero modes, which are related to Gribov copies, will be ignored in this paper. The Hodge-DeRham decomposition of any *real-*valued form $A$ follows from

$$1 = \delta \Delta^{-1} d + d \Delta^{-1} \delta \qquad (14)$$

which implies $A = \delta \Delta^{-1} F + d\omega$ where $dF = 0 = \delta\omega$. For *integer* $j$-form $l$, the Hodge-DeRham decomposition takes the slightly modified form [13]

$$l(c_j) = \delta \Delta^{-1} v(c_{j+1}) + d\Delta^{-1} r(c_{j-1}), \quad v = dl, \quad dv = 0, \quad \delta r = 0. \qquad (15)$$

Then one can show [14] that up to an overall volume proportionality factor

$$\sum_{l(c_j) \in \mathbb{Z}} \propto \sum_{\{v(c_{j+1}) \in \mathbb{Z} | dv = 0\}} \sum_{\{r(c_{j-1}) \in \mathbb{Z} | \delta r = 0\}}. \qquad (16)$$

Hodge-DeRham decomposition of $A$ and $l$, application of (16), and absorption of all the exact pieces into $d\varphi$ yields

$$\Delta_{FP}^{-1}(A) = \mathcal{A}(A) \; \mathcal{Z}_v(A) \int_{-\infty}^{\infty} \mathcal{D}\varphi \; \exp\left\{ -\frac{\kappa}{2}(d\varphi, d\varphi) + \kappa(\varphi, \delta A) \right\} \qquad (17)$$

where $\mathcal{A}(A) \equiv \exp\left\{-\frac{\kappa}{2}\|A\|^2\right\}$ and

$$\mathcal{Z}_v(A) \equiv \sum_{\{v(c_2) \in \mathbb{Z} | dv = 0\}} \exp\left\{-2\pi^2 \kappa(v, \Delta^{-1}v) + 2\pi\kappa(v, \Delta^{-1}dA)\right\}. \qquad (18)$$

In the RHS of (17) the overall proportionality constant arising from (16), which has no dynamical consequences, has been dropped.

Due to the condition $\delta^* v = 0 (dv = 0)$, the vortices $^*v$ in $\mathcal{Z}_v$ are *closed* Abrikosov strings in $D = 2 + 1$ and *closed* Nielsen-Olesen string worldsheets in $D = 3 + 1$. Due to the factorization of $\mathcal{A}$ and $Z_v$ from the $\varphi$ integral, electromagnetic sources from $\mathcal{A}$ and vortices from $\mathcal{Z}_v$ are *quenched* in $\mathcal{Z}[\mathcal{O}]$, that is, $\mathcal{A}$ and $\mathcal{Z}_v$ drop out of $G(A)$ in Eqs. (9-10). Putting all the pieces together yields

$$\begin{aligned}
G(A) &= \frac{\sum_{r(c_0) \in \mathbb{Z}} \exp\left\{-2\pi^2 \kappa(r, \Delta^{-1} r) + 2\pi\kappa(\Delta^{-1} r, \delta A)\right\}}{\int_{-\infty}^{\infty} \mathcal{D}\varphi \; \exp\left\{-\frac{\kappa}{2}(d\varphi, d\varphi) + \kappa(\varphi, \delta A)\right\}} \\
&= \frac{\sum_{r(c_0) \in \mathbb{Z}} \exp\left\{-2\pi^2 \kappa((r - \frac{\delta A}{2\pi}), \Delta^{-1}(r - \frac{\delta A}{2\pi}))) + \frac{\kappa}{2}(\delta A, \Delta^{-1} \delta A)\right\}}{\exp\left\{\frac{\kappa}{2}(\delta A, \Delta^{-1} \delta A)\right\}} \\
&= \sum_{r(c_0) \in \mathbb{Z}} \exp\left\{-2\pi^2 \kappa \left((r - \frac{1}{2\pi}\delta A), \Delta^{-1}(r - \frac{1}{2\pi}\delta A)\right)\right\} \quad (19)
\end{aligned}$$

where $\delta r(c_0) = 0$ is automatic since $r$ is a 0-form. Taking the $\kappa \to \infty$ limit yields the lattice Landau gauge gauge fixing condition

$$\lim_{\kappa \to \infty} G(A) = \sum_{r(c_0) \in \mathbb{Z}} \delta(\delta A - 2\pi r). \quad (20)$$

Eq. (20) implies that $\delta A$ is ambiguous up to $2\pi r$ where $r(c_0) \in \mathbb{Z}$, $\delta r = 0$. These integer gauge ambiguities do not affect Dirac strings or any operators which are periodic in $A_\mu$.

## 4 Origin of Photon Mass

The photon mass contribution from the gauge fixing action and the Faddeev-Popov determinant is given by the $\|A\|^2$ term in the functional Taylor expansion of $\log G(A)$. Since $\log G$ depends only on the longitudinal part of $A$, it cannot have an $\|A\|^2$ term in its Taylor expansion and, hence, gauge fixing operator $G(A)$ cannot generate a nonzero photon propagator mass pole. The photon mass must come from $S_c$.

To isolate where the nonzero $M_\gamma$ comes from, introduce the noncompact QED action

$$S_{nc}(A) \equiv \frac{\beta}{2} \|dA\|^2 \quad (21)$$

where $A \in (-\infty, \infty)$. $S_{nc}$ is just the lattice transcription of massless continuum QED whose Landau gauge photon propagator is massless. Expression (8) can be rewritten as

$$\mathcal{Z}[\mathcal{O}] = \int \mathcal{D}A \ F(A) \ \exp^{-S_{nc}(A)} \ G(A) \ \mathcal{O}(A), \qquad (22)$$

$$F(A) \equiv \exp^{-S_c(A)} \exp^{+S_{nc}(A)} . \qquad (23)$$

Since $S_{nc}$ and $\log G$ have no photon mass contribution, the photon propagator pole is determined completely by

$$\log F(A) = -\frac{1}{2} M_\gamma^2 \ \|A\|^2 \ + \ \cdots \qquad (24)$$

To probe into the origin of $M_\gamma$, make the variables change

$$N = \delta \Delta^{-1} m + \mathrm{d} \Delta^{-1} q, \quad \mathrm{d}m = 0, \quad \delta q = 0 \qquad (25)$$

on $N$ of Eq. (11). $(D-3)$-form $^*m$ is the monopole current and, since

$$\delta^* N = {}^*m, \qquad (26)$$

$^*N$ is the Dirac "sheet" in $D = 3+1$ dimensions. Application of Eq. (16) yields

$$F(A) = C_m \sum_{\{q(c_1) \in \mathbb{Z} | \delta q = 0\}} \exp\left\{-2\pi^2 \beta(q, \Delta^{-1} q) + 2\pi \beta(A, q)\right\} \qquad (27)$$

where

$$C_m \equiv \sum_{\{m(c_3) \in \mathbb{Z} | dm = 0\}} \exp\left\{-2\pi^2 \beta(m, \Delta^{-1} m)\right\} . \qquad (28)$$

Since monopole prefactor $C_m$ is independent of $A$, we have immediately that monopoles do not contribute to the photon propagator pole. We conclude, in agreement with the numerical results of Section 2, that the interaction of the integer valued 1-form $q$ and the gauge field $A$ generate the photon mass. In other words, as it follows from Eq. (25), the Dirac strings "without" monopoles are responsible for the photon mass.

# 5 Acknowledgments

MIP thanks the LSU Physics Department for its hospitality. MIP and KY are indebted to Dick Haymaker and Lai Him Chan for providing a pleasant working environment, and KY thanks Claude Bernard and Amarjit Soni for the use of their NERSC(Livermore) account for the nonabelian work [2] closely related to this project. Analysis and the Figures were done at the LSU Concurrent Computing Lab and, also, on the NERSC supercomputer time grant of KY. MIP has been partially supported by a grant of the American Physical Society. KY is supported by U.S. DOE grant DE-FG05-91ER40617.

# Appendix A Landau and cLandau Gauge

Gauge-fixing is determined by extremizing with respect to gauge transformations $V_x = \exp\{-i\phi(x)\}$ the action [5, 6]

$$S^L \equiv \sum_x S^L_x, \quad S^L_x \equiv -\sum_{\mu=1}^{D} \cos(A_\mu(x) - \phi(x+\hat{\mu}) + \phi(x)). \tag{A.1}$$

At a site $S^L_x$ is extremized exactly by choosing $\phi(x) = -\arctan(a/b)$ where

$$a(x) \equiv \sum_{\mu=1}^{D} \sin A_\mu(x) - \sin A_\mu(x - \hat{\mu}), \tag{A.2}$$

$$b(x) \equiv \sum_{\mu=1}^{D} \cos A_\mu(x) + \cos A_\mu(x - \hat{\mu}). \tag{A.3}$$

Arctan is multivalued. Landau gauge is given by the branch choice

$$\text{Landau gauge}: \quad \phi \in \begin{cases} [-\frac{\pi}{2}, \frac{\pi}{2}) & b > 0; \\ [\frac{\pi}{2}, \pi) & b < 0, a < 0; \\ [-\pi, -\frac{\pi}{2}) & b < 0, a > 0; \end{cases} \tag{A.4}$$

which *minimizes* $S^L_x$. This branch choice requires simultaneously that $a(x) = 0$ and $A_\mu \to 0$ in the $\beta \to \infty$ limit, and agrees with the Landau gauge of Ref. [6].

The Landau gauge used in Ref. [12] is not this but an inequivalent extension of continuum Landau gauge. To distinguish it from the Landau gauge in this paper and Ref. [6], we call it "cLandau" gauge in this paper. The branch choice for cLandau gauge is

$$\text{cLandau gauge}: \quad \phi \in [-\frac{\pi}{2}, \frac{\pi}{2}) \; \forall\, a, b. \tag{A.5}$$

This branch choice also converges to $a(x) = 0$ but, unlike Landau gauge, preserves the discrete global "compactness" symmetry

$$A_\mu \to \begin{cases} \pi - A_\mu & \text{if } A_\mu > 0; \\ -(\pi + A_\mu) & \text{if } A_\mu \leq 0. \end{cases} \tag{A.6}$$

In the $\beta \to \infty$ limit of cQED in cLandau gauge, half of the $A_\mu$ tends to 0, a quarter tends to $\pi$ and the remaining quarter to $-\pi$.